%% file: LOOPS_34_NoRH_arXiv.tex
\documentclass[twocolumn]{aastex631}
\def\gsim{\ \raise 3pt \hbox{$\rangle$} \kern -8.5pt \raise -2pt \hbox{$\sim$}\ }
\usepackage{graphicx,natbib,color,amsmath,subfigure}
\usepackage{longtable}
\usepackage{ulem}
\newcommand{\blank}[1]{}
\usepackage{color}

\begin{document}

\include{definitions}

\title{Giant post-flare loops in active regions with extremely strong coronal magnetic fields}

\author[0000-0003-1358-5522]{Costas E. Alissandrakis}
\affil{Deparment of Physics, University of Ioannina, Ioannina 45110, Greece}
\author[0000-0001-5557-2100]{Gregory D. Fleishman}
\affil{Center For Solar-Terrestrial Research, New Jersey Institute of Technology, Newark, NJ 07102}
\affil{Institut f\"ur Sonnenphysik (KIS), Sch\"oneckstrasse 6, D-79104 Freiburg, Germany}
\author[0000-0003-2990-1390]{Viktor V. Fedenev}
\affil{Institute of Solar-Terrestrial Physics SB RAS, Lermontov st., 126a, Irkutsk, 664033 Russia}
\author[0000-0002-8574-8629]{Stephen M. White}
\affil{Air Force Research Laboratory, Space Vehicles Directorate, Kirtland AFB, NM 87123, USA}
\author[0000-0002-1589-556X]{Alexander T. Altyntsev}
\affil{Institute of Solar-Terrestrial Physics SB RAS, Lermontov st., 126a, Irkutsk, 664033 Russia}

\begin{abstract}
We report for the first time the detection of thermal free-free emission from post-flare loops at 34\,GHz in images from the Nobeyama Radioheliograph (NoRH). We studied 8 loops, 7 of which were from regions with extremely strong coronal magnetic field reported by \cite{2023ApJ...943..160F}. Loop emission was observed in a wide range of wavelength bands, up to soft X-rays, confirming their multi-temperature structure and was associated with noise storm emission in metric $\lambda$. The comparison of the 17\,GHz emission with that at 34\,GHz, after a calibration correction of the latter, showed that the emission was optically thin at both frequencies. We describe the structure and evolution of the loops and we computed their density, obtaining values for the top of the loops between 1 and $6\times10^{10}$\,cm$^{-3}$, noticeably varying  from one loop to another and in the course of the evolution of the same loop system; these values have only a weak dependence on the assumed temperature, $2\times10^6$\,K in our case, as we are in the optically thin regime. Our density values are above those reported from EUV observations, which go up to about $10^{10}$\,cm$^{-3}$. This difference  could be due to the fact that different emitting regions are sampled in the two domains and/or due to the more accurate diagnostics in the radio range, which do not suffer from inherent uncertainties arising from abundances and non-LTE excitation/ionization equilibria. We also estimated the magnetic field in the loop tops to be in the range of 10-30\,G. 
\end{abstract}

\keywords{Sun: radio radiation---Sun: corona---Sun: flares---Sun: magnetic fields}

\section{Introduction}\label{sect:intro}

Loops are a common structure in the solar corona, mapping closed magnetic field configurations. Post-flare loops (loop prominences), seen in eclipses and in strong chromospheric and coronal lines of the optical range, have been known for a long time \citep[see, e.g.,][]{1963sofl.book.....S, 1969soat.book.....Z, 1976sofl.book.....S}.  It was in the 70's, thanks to the high spatial resolution X-ray images from Skylab, that loops were recognized as the most abundant structural element of the corona, both in active regions and in the quiet Sun outside coronal holes. Related works are too numerous to cite them all, so we mention here the monographs of \cite{1991plsc.book.....B}, \cite{2005psci.book.....A}, \cite{2019ASSL..458.....A}, \cite{2018sflo.book.....H} and the review of \cite{2014LRSP...11....4R}.

Loops appear in a variety of sizes, temperatures and densities. On the basis of their temperature, they are often classified in three groups: cool loops, seen in wavelength regions where the radiation forms below $10^5$\,K, warm loops formed in the temperature range of $10^5$\,K to $10^6$\,K and hot loops formed above $10^6$\,K. This classification does not reflect their origin and size, which range from tiny loops associated with X-ray bright points, through flare and post-flare loops, medium scale active region loops, to large scale quiescent loops.

Post-flare loops appear in conjunction with large flares. Among their basic properties are their duration of several hours, downflows from their top to the footpoints and their apparent expansion. The latter is well seen beyond the limb in \ha, with the tops of the loops reaching heights of 100\,Mm or more; it is due to the formation of new, higher loops, and the fading of the old ones, as the reconnection region  above the cusp of the loops moves higher. Post-flare loops are also visible in wavelength bands forming in transition region and coronal temperatures, revealing their multi-temperature nature. In recent years white light emission from post-flare loop has been detected \citep{2014ApJ...780L..28M,2018ApJ...867..134J,2023A&A...672A..32F} in HMI images. 

Loop densities and temperatures are of particular relevance to the present article. Measurements of cool post-flare loops in the optical range, compiled by \cite{1991plsc.book.....B} in their table 2.9, give temperatures, $T_e$, in the range of $7\times10^3$\,K to $25\times10^3$\,K and densities, $N_e$, from $3\times10^9$\,cm$^{-3}$ to $6\times10^{12}$\,cm$^{-3}$. Going to higher temperatures, \cite{1978ApJ...223.1046F} measured a number of active region loops, including post-flare loops, in Skylab EUV images formed between $0.2\times10^6$\,K to $2\times10^6$\,K, and deduced $N_e$ in the range of  $0.3\times10^9$\,cm$^{-3}$ to $5.3\times10^9$\,cm$^{-3}$, while \cite{1980SoPh...65..347C} analyzed Skylab spectra of EUV lines formed in the $T_e$ range of $0.4\times10^6$\,K to $4\times10^6$\,K and reported densities from $1.3\times10^9$\,cm$^{-3}$ to $10^{10}$\,cm$^{-3}$ for a loop prominence seen at the limb. Using Yohkoh/SXT images, \cite{1995SoPh..156..337S} derived an electron density of $\simeq7\times10^9$\,cm$^{-3}$ for hot post-flare loops at $T_e\simeq5.5\times10^6$\,K. \cite{2000A&A...355..769V} studied decaying post-flare loops from SOHO/CDS and Yohkoh/SXT data; from the SXT images they deduced $N_e$ values of $\sim1.5\times10^{10}$\,cm$^{-3}$ at the beginning of their data and $\sim6\times10^{9}$\,cm$^{-3}$ 16\,min later, while the ratio of the density sensitive Fe{\sc xiv} 334.2\,AA/353.8\,\AA\ lines from CDS near the latter time interval gave densities in the range of $7.25\times10^{9}$\,cm$^{-3}$ to $9.9\times10^{9}$\,cm$^{-3}$. Overall, densities of up to about $10^{10}$\,cm$^{-3}$ have been reported for high-temperature post-flare loops, varying significantly during the decay phase. The temperature of white light loops cannot be determined, still \cite{2018ApJ...867..134J} estimated densities from $2.2\times10^{12}$\,cm$^{-3}$ to $7.3\times10^{12}$\,cm$^{-3}$ for cool (6000\,K) loops and from $4.5\times10^{12}$\,cm$^{-3}$ to $1.5\times10^{13}$\,cm$^{-3}$ for hot ($10^6$\,K) loops.

Quiescent loops have lower density. \cite{1999ApJ...517L.155L} analyzed TRACE observations of 4 limb loops and deduced temperatures in the range of 1\,MK to 1.3\,MK and emission measures from 0.8 to $6\times10^{27}$\,cm$^{-5}$ near the base of the loops.  \cite{2003A&A...406.1089D} used TRACE and SOHO/CDS observations of a limb loop and obtained $\log T$ of 5.83 at the base of the loop and 5.95 at a distance of 60\arcsec, density of 38 and 2.5$\times10^8$\,cm$^{-3}$ and emission measure of 19 and $0.5\times10^{26}$\,cm$^{-5}$, respectively.  \cite{2008ApJ...680.1477A} deduced mean loop temperatures of $1.1\pm0.2$\,MK and loop base densities of  $(2.2\pm0.5)\times10^9$\,cm$^{-3}$ from the analysis of 30 loops observed with EUVI on board STEREO A and B; the loop lengths were $130\pm 67$\,Mm and their widths $1.1\pm0.3$\,Mm, while the density profiles were found hydrostatic with scale height $\lambda_T(T_e)=47 (T_e/1\mbox{ MK})$\,Mm. \cite{2011ApJ...730...85B} studied 3 active region fan loops with SDO/AIA and Hinode/EIS and reported $\log T$ in the range of  5.88 to 6.05 and $\log$ emission measure in the range of 26.10 to 27.18. These results converge towards nearly isothermal loops at $\sim10^6$\,K, with a density around $3\times10^9$\,cm$^{-3}$ and emission measure of about $10^{27}$\,cm$^{-5}$  near their base.

In the microwave range there are many observations of flare loops \citep[see references in ][]{2018sflo.book.....H}, but few of post-flare loops \citep[e.g.][]{1981ApJ...243L.103V, 1983SoPh...83....3S, 1984ApJ...279..427W}. \cite{1994kofu.symp..181H, 1999spro.proc..153H} presented a nice image of post-flare loops at 17\,GHz, also shown on the cover of the Kofu symposium proceedings \citep{1994kofu.symp......}, and compared it with SXT images. 

In a recent article, \cite{2023ApJ...943..160F} reported high brightness temperatures in sunspot-associated 34\,GHz sources, which they associated with thermal gyroresonance (GR) emission in the presence of very strong magnetic fields. While examining the images of the associated active regions, we detected long-lived 34\,GHz emission from large-scale loop systems, which were also present in TRACE images in the 171\,\AA\ and 195\,\AA\ bands. The examination of the X-ray flux from GOES and soft X-ray images from GOES/SXI revealed that these were post-flare loops. To the best of our knowledge, such emissions have not been reported before, although \cite{2003ApJ...595L.111W} mentioned the presence of a large, hot, long-lived loop at 34\,GHz, not visible at other wavelengths, observed before an X-class flare. These observations present a unique opportunity to study loops in a wavelength range which is not plagued with uncertainties such as excitation and ionization equilibria and elemental abundances. 

\begin{figure*}
\includegraphics[width=.995\textwidth]{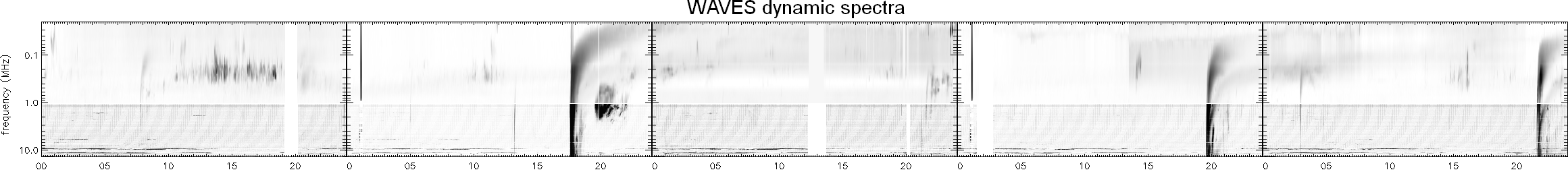}
\includegraphics[width=\textwidth]{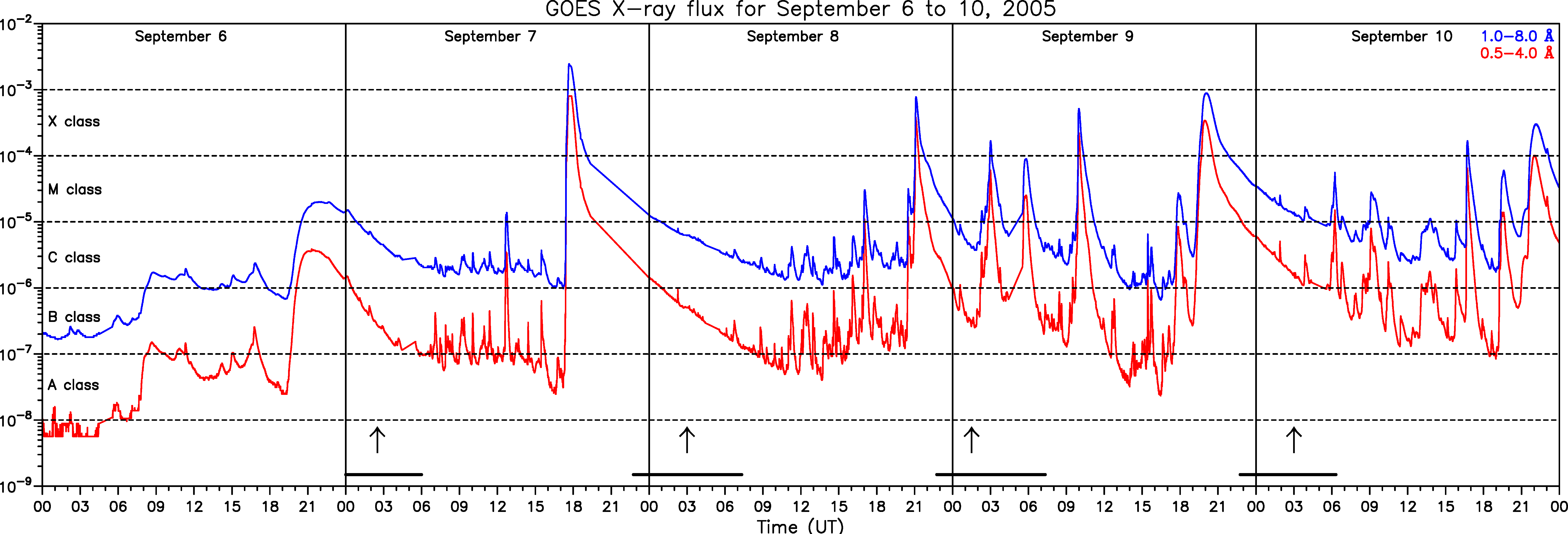}
\includegraphics[width=.995\textwidth]{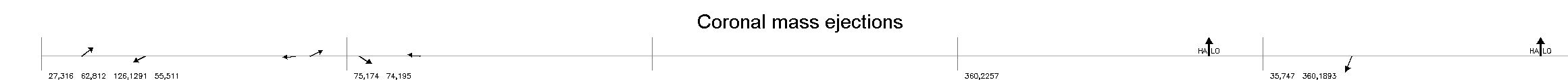}
\caption{Time profiles of emissions from AR 10808 during September 6 to 10, 2005. Top: Dynamic spectrum from WAVES in the frequency range of 15 to 0.02\,MHz. Middle: GOES X-ray flux in the 1.0-8\,\AA\ and 0.5-4\,\AA\ channels. The horizontal bars above the time axis give the intervals of NoRH observations and the arrows mark the times of the images displayed in Figure~\ref{fig:overview}.  Bottom: CMEs; the arrows show the direction of each CME and the numbers give its width and velocity. The WAVES and CME data are composite images from https://secchirh.obspm.fr/}
\label{fig:GOES_WAVES_CME}
\end{figure*}

In this work we present and analyze observations of post-flare loops, most of them from active regions included in the work of  \cite{2023ApJ...943..160F}, with emphasis on the emission mechanism and the information about their physical conditions that can be provided by radio observations in the mm-$\lambda$ range. After this introduction, we present the observations (Section~\ref{sect:obs}). In Section~\ref{sect:10808} we analyze in detail loops associated with AR 10808: in Section~\ref{sect:overview} we give an overview of the data, in Section~\ref{sect:NRH_thermal} we discuss metric-$\lambda$ emission and in Section~\ref{sect:evolution} we present the evolution of the loops. In Section~\ref{sect:other} we briefly discuss 4 more loop systems from other active regions, while in Section~5 we present the analysis of our data and our results: we discuss the optical thickness of the NoRH emission (5.2.1), the range of electron temperature derived from NoRH and SSRT observations (5.2.2) and our estimates the loop physical parameters (5.2.3). Finally, in Section~\ref{sect:conclude} we summarize results and in Section\ref{sect:discuss} we present our conclusions.

\section{Observations}\label{sect:obs}
The main source of data used in this study was the Nobeyama Radioheliograph (NoRH) full-disk microwave imaging observations of the Sun \citep{1994IEEEP..82..705N} at 17\,GHz (Stokes parameters $I$ and $V$) and 34\,GHz (Stokes $I$ only). NoRH data include radio images, correlation curves, as well as raw visibilities. The data are available for every observing day from 22:45\,UT to 06:30\,UT regardless the time of year. The cadence is 1 second for the correlation curves  and for the raw  data \citep{1994IEEEP..82..705N}. The nominal spatial resolution is 10\arcsec\ at 17\,GHz and 5\arcsec\ at 34\,GHz. The actual resolution is worse than that; from measurements of point-like sources and of the width of the limb profile, we obtained values larger by a factor of two. In addition to the NoRH images, we used full-disk images at 5.7\,GHz from the Siberian Solar Radio Telescope (SSRT) with $\sim30$\arcsec\ resolution.

Visibility data are available  at the NoRH site with a cadence of 1\,s, as well as pre-synthesized images with a 10-minute cadence for 17\,GHz, and daily images for 34\,GHz. The radio images used in this work were synthesized as in \citet{2023ApJ...943..160F} from the raw visibility data using the Hanaoka synthesis code following classical CLEAN algorithm \citep{1974A&AS...15..417H}  available from NoRH package in SolarSoftWare \citep[SSW, ][]{1998SoPh..182..497F} library of programming language IDL (Interactive Data Language). We produced images throughout the day at a cadence of 30\,min with an integration time of 100\,s, in order to increase the dynamic range. Raw datasets were automatically downloaded from the NoRH database by the package routines.

\begin{figure*}
\includegraphics[width=\textwidth]{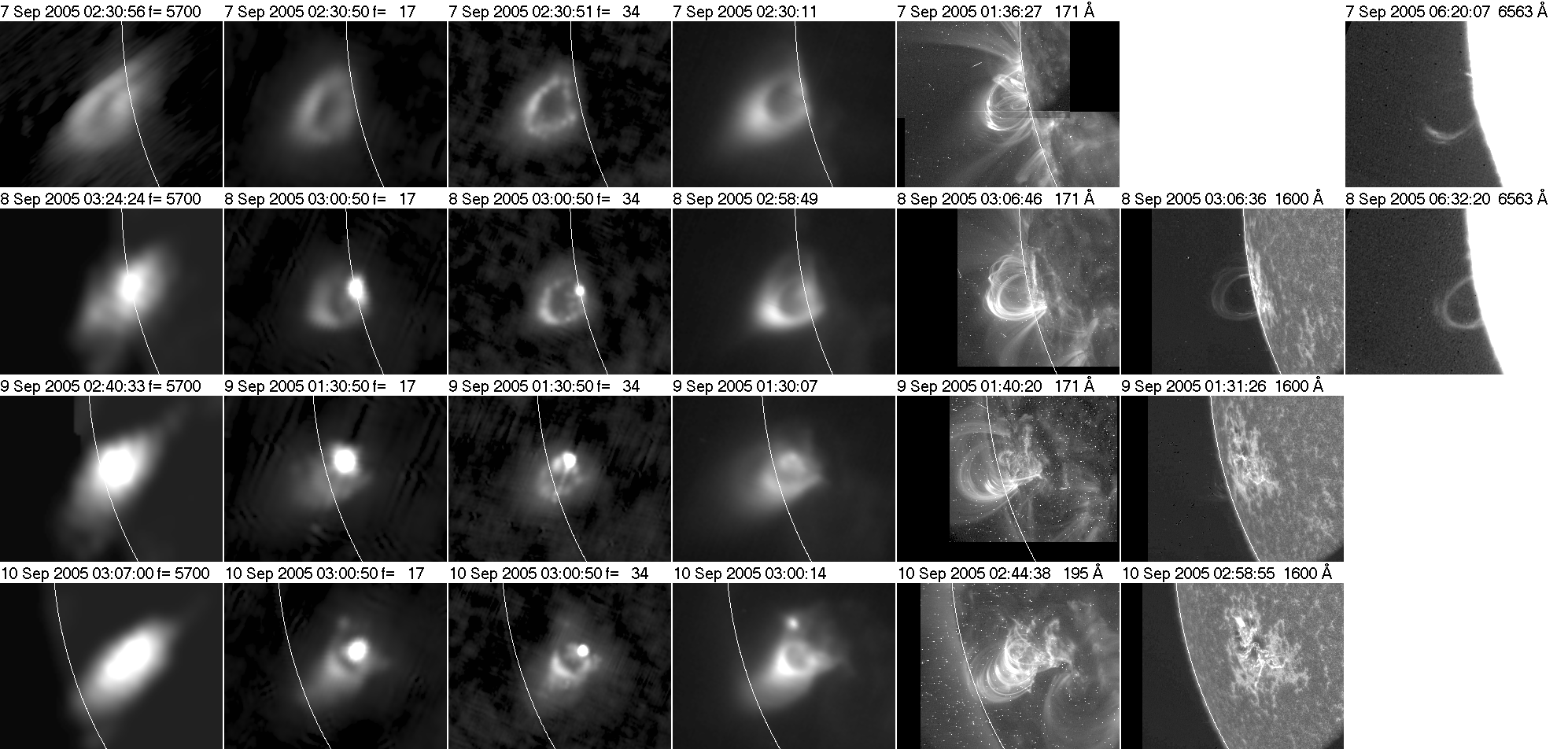}
\caption{Daily images of the loop system in AR 10808 from September 7 to September 10, 2005, in a 510\arcsec\ by 380\arcsec\ field of view. From left to right: SSRT at 5.7\,GHz, NoRH at 17\,GHz, NoRH at 34\,GHz, SXI in soft X-rays, TRACE in the 171\,\AA\ or 194\,\AA\ band, TRACE in the 1600\,\AA\ band and Kanzelhoehe \ha. The emission from the solar disk has been subtracted from the radio images. The white arc marks the photospheric limb.}
\label{fig:overview}
\end{figure*}

Accurate absolute $T_{\sun}$ calibration is very important for our analysis. The SSRT images are calibrated under the assumption that the quiet Sun brightness temperature at the center of the disk, $T_{b0}$, is $16\times10^3$\,K, derived from the measurements of \cite{1991ApJ...370..779Z}, which are considered to be the most reliable to date. The calibration of the NoRH images at 17\,GHz is also based on the measurements of \cite{1991ApJ...370..779Z}, which go up to 18\,GHz and give $T_{b0}=10.6\times10^3$\,K at 17\,GHz. The same value is also adopted by the NoRH software for the 34\,GHz images, which, obviously, is  incorrect; hence a re-calibration is required. 

To that end we used the information obtained from the analysis of the center-to-limb variation (CLV) of the brightness in ALMA full-disk images by \cite{2023A&A...670C...5A} to obtain $T_{b0}=8378$\,K at 34\,GHz. This value is consistent with the compilation of $T_b$ measurements in the mm-$\lambda$ range by \cite{2004A&A...419..747L}, re-calibrated by these authors according to an accurate reference spectrum of  the moon; indeed, a quadratic fit of the $T_{b0}$ values between 1 and 21 mm gives $T_{b0}=8458$\,K at 34\,GHz. These values give a correction factor of 0.80 for the NoRH images at 34\,GHz, which we adopted in this work. We note that a comparison of the CLV curves at 17 and 34\,GHz with the method of \cite{2023A&A...670C...5A} for the images of September 7, 2005, gave a slightly higher correction of 0.86; we preferred not to use that, because it is based on measurements during a single day, which is prone to an uncertainty.

Additional data include UV and EUV images from the Transition Region and Coronal Explorer (TRACE) with $\sim0.5$\arcsec\ resolution, EUV images from SoHO/EIT, soft X-ray images from GOES/SXI and Yohkoh/SXT, metric wavelength images from the Nan\c cay  Radioheliograph (NRH) and \ha\ images from the Kanzelhoehe solar observatory. For the time evolution of the flux we used soft X-ray data in the 1-8\,\AA\ and 0.5-4\,\AA\ bands from GOES-12 and dynamic spectra from the WAVES instrument aboard the WIND satellite in the decametric-hectometric wavelength range.

\medskip
\section{Loops associated with AR 10808}\label{sect:10808}
\subsection{Overview}\label{sect:overview}

A global view of the active region evolution is provided by the GOES measurements of the X-ray flux.  The corresponding plot for our period of interest is given in the middle panel of Figure~\ref{fig:GOES_WAVES_CME} and reveals that AR 10808 was very active, with several X-class flares. Indicative of the magnitude of these flares is the fact that we found white-light emission in the TRACE images at 19:37\,UT and 20:06\,UT on September 9. Some Coronal Mass Ejections were detected (bottom panel of the figure) and 3 of the big flares gave emissions in the decametric-hectometric wavelength range, recorded by the WIND/WAVES instrument  (top panel of Figure~\ref{fig:GOES_WAVES_CME}).

\begin{figure*}
\begin{center}
\includegraphics[width=\textwidth]{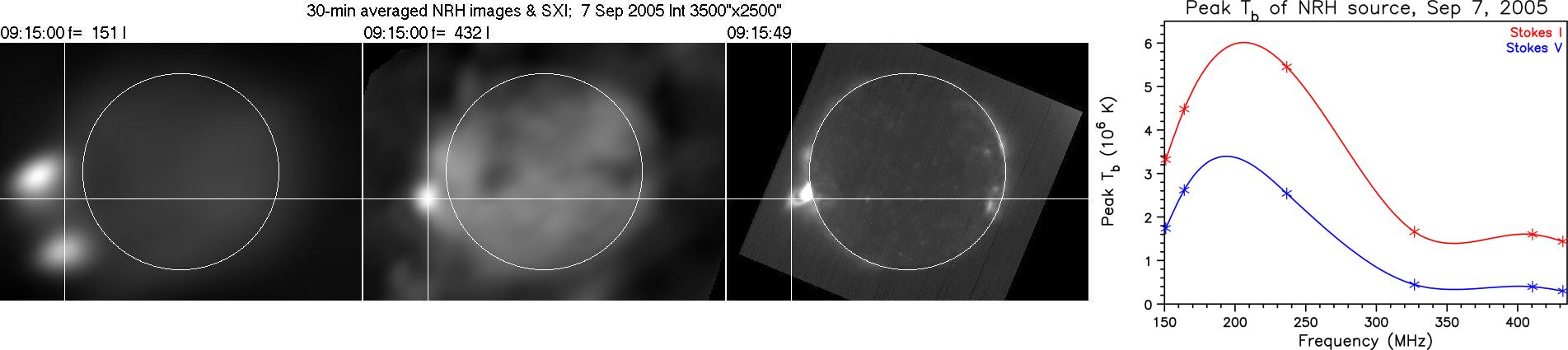}
\caption{Left: NRH images at 150.9 and 432\,MHz, together with the corresponding SXI image (overexposed). The cursor is at the peak of the 432\,MHz source and the white circle marks the photospheric limb. Right: Peak brightness temperature spectrum of the NRH images (150.9, 164, 236.6, 327, 410.5 and 432 MHz), in total intensity (Stokes $I$) and circular polarization (Stokes $V$). The solid lines are spline interpolations through the measured points (asterisks).}
\label{fig:NRH_SXI}
\end{center}
\end{figure*}

Figure~\ref{fig:overview} shows an overview of the loop system of AR 10808 from September 7 to September 10, 2005. The active region itself was large, with a complex magnetic field configuration \citep[see figure 12 in][]{2023ApJ...943..160F}. For September 8 and 10, images near the NoRH local meridian crossing time are displayed, for September 7 images near the SSRT image are shown, whereas for September 9 images at a time when the AR emission was not very bright were selected. \ha\ images are 3 to 5 hours later than the NoRH images. All images of each column are displayed with the same color table, up to $10^6$\,K, $5\times10^5$\,K and $10^5$\,K for the 5.7\,GHz, 17\,GHz and 34\,GHz images, respectively; as a result the strong sunspot-associated source in the radio images of September 8 to 10 is saturated.

On September 10 the TRACE EUV images show a long arcade of large scale loops; this arcade is also detectable in the NoRH and SXI images of the same day, in spite of the lower resolution. It is possible that similar arcades existed during the previous days, partly or fully concealed due to projection effects, as the region was closer to the limb. Projection effects also decreased the apparent height of the loops, which was maximal on September 7, when the active region as well as the loop footpoints were just behind the limb. The projected height of the loop on that day was about 85\, Mm and the distance between its footpoints about 60 Mm. Interestingly, the SXI images show evidence of a cusp structure above the loop top.

The NoRH loops were strongest on September 10, with an observed peak brightness of $2.2\times10^5$\,K at 17\,GHz and $8.1\times10^4$\,K at 34\,GHz. The maximum was near the top of the loop, where we probably have a superposition of many unresolved loops. The 17\,GHz NoRH images of September 7 showed a weak circular polarization signal of $\sim0.4$\%; although the NoRH might not be able to reliably measure such a low polarization, this value could put an upper limit of $\sim12$\,G to the longitudinal magnetic field of the loop. The magnetic field was evaluated from the polarization, $P$, using the expression \citep[cf.] []{2021FrASS...7...77A}:
\begin{equation}
P=n\frac{2.8\times10^6B_\|}{f}   
\label{Eq:Pol_ff}
\end{equation}
where $n=\ln T_b/\ln f$ is the spectral index (2 for a uniform optically thin source). 

We note that there was no measurable polarization on the loop during the other days, apparently because of the presence of the bright sunspot-associated source and the limited dynamic range of the images. The 5.7\,GHz SSRT image on September 7 had a maximum of $5\times10^5$\,K. Near the top of the loop the circular polarization was $\sim4$\%, which leads to a longitudinal magnetic field estimate of $\sim40$\,G, a more accurate value than the one from the NoRH due to the longer observing wavelength of the SSRT.

The available 1600\,\AA\ and \ha\ images show emission from low-temperature plasma which became very weak on September 9, disappeared on September 10 and showed no trace on the solar disk. We thus have a loop system with a large range of temperatures, from chromospheric to coronal; as there are structural differences among the images of the same day in wavelength regions with different response to the plasma temperature, we may conclude that the high and low temperature loops are discrete. As the observed brightness temperature in the radio images is well above that of the chromosphere, the microwave emission, if thermal, comes from the high temperature loops.

\subsection{Emission at metric wavelengths}\label{sect:NRH_thermal}
In the metric wavelength range, NRH observations revealed the presence of noise storm emission above AR 10808, which was visible from September 4 2005, 4 days before the rising of the AR from the east limb, until September 19, 2005, 3 days before its setting in the west limb. 

Figure~\ref{fig:NRH_SXI} shows the situation near the start of NRH observations, at 09:15 UT of September 7, 2005, with images at the lowest (150.9\,MHz) and highest (432\,MHz) NRH frequencies as well as the corresponding SXI image. Although the 150.9\,MHz sources are clearly non-thermal, with brightness temperature of $\sim5\times10^6$\,K and circular polarization of about 50\%, the sources above 350\,MHz could be thermal, as $T_b$ dropped to $\sim 1.5\times10^6$\,K and the polarization to around 20\% (see the plot of the spectrum in Figure~\ref{fig:NRH_SXI}, right). Moreover, the position of the peak of the 432\,MHz source, marked by the cursor,
shifted near the top of the SXI loop; the cursor is about 150\,Mm above the photosphere, reflecting the expansion of the loop in the time elapsed since the images of Figure~\ref{fig:overview}. 

If the NRH emission at 432\,MHz is indeed thermal, it is tempting to consider the possibility that it might be associated with the top of the loop visible in other wavelength ranges. However, we should bear in mind that scattering will have decreased the brightness and refraction will have shifted the source towards the disk \citep[e.g.][]{2019ApJ...884..122K}. In view of these issues, we prefer not to examine this possibility, except for noting that it would imply an upper limit of $2.32\times10^8$\,cm$^{-3}$ for the density of the emitting region, which is the electron density corresponding to a plasma frequency of 432\,MHz.

\begin{figure}[h]
\begin{interactive}{animation}{Movie1.mp4}
\centerline{\includegraphics[width=.44\textwidth]{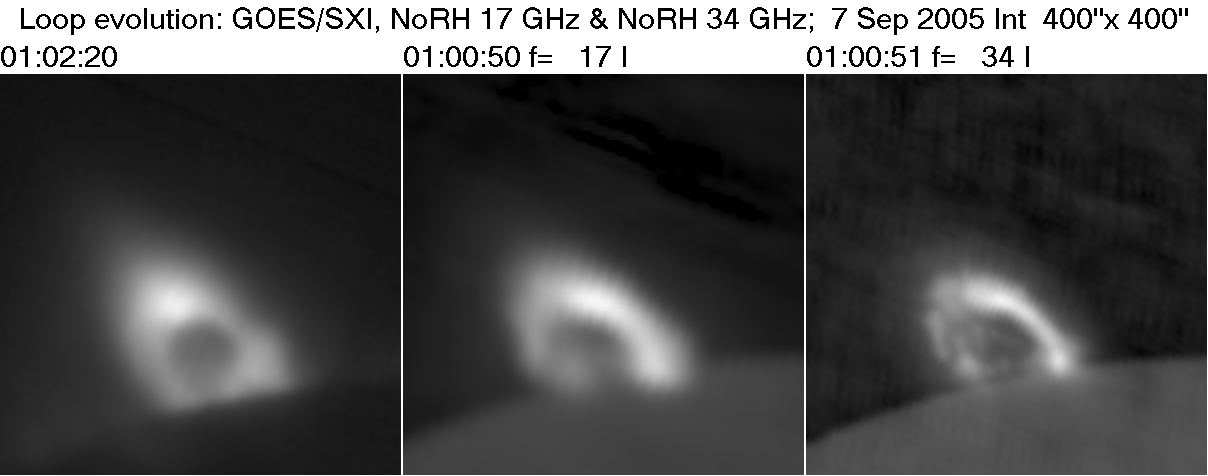}}
\end{interactive}
\caption{Loop images from GOES/SXI (left), NoRH 17\,GHz I (middle) and NoRH 34\,GHz I (right) for the system observed on September 7, 2005. Solar east is up and the field of view is 400\arcsec\ by 400\arcsec. An animation of these images is available, covering the time interval 00 -- 06 UT of September 7, 2005. The range of display values and the contrast is the same for all images of the same wavelength, with a maximum of 240\,kK for 17\,GHz and 90\,kK for 34\,GHz.}
\label{fig:EvolNoRH7}
\end{figure}

\subsection{Time evolution}\label{sect:evolution}

The NoRH observing intervals coincided with the decay of large flares. It is thus obvious that the loops detected at 17 and 34\,GHz are post-flare loops, associated with the M1.4 occulted flare of September 6 (SOL2005-09-06T19:32:00), the X17 flare of September 7 (SOL2005-09-07T17:17:00), the X5.4 flare of September 8 (SOL2005-09-08T20:52:00) and the X6.2 flare of September 9 (SOL2005-09-09T19:13:00), all of which were long duration events (LDE), lasting several hours. We note that, at least for the occulted flare of September 6, practically all the X-ray flux recorded by GOES came from the post-flare loops, as there were no active regions on the solar disk. 

\begin{figure}
\begin{interactive}{animation}{Movie2.mp4}
\centerline{\includegraphics[width=.44\textwidth]{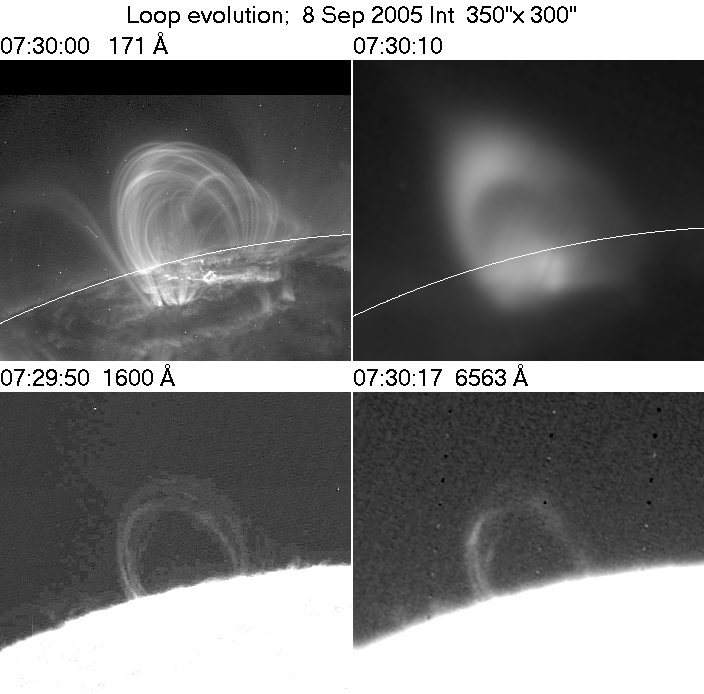}}
\end{interactive}
\caption{Images of the loop system on September 8, 2005 in TRACE 171\,\AA\ (top left), GOES/SXI (top right), TRACE 1600\,\AA\ (bottom left), and Kanzelhoehe \ha\ (bottom right). Solar east is up. The white arc marks the photospheric limb and the field of view is 350\arcsec\ by 300\arcsec. An animation of these images is availabe, from 02:40 to 14:40 UT of September 8, 2005.}
\label{fig:EvolSep8}
\end{figure}

The loops were highly variable in time, both in intensity and in shape, as shown in Figure~\ref{fig:EvolNoRH7}, which displays our set of NoRH images for September 7, together with the corresponding SXI images. The loop top appears to rise slowly with an apparent speed of $\sim1.4$\,km/s, derived from the NoRH images. As SXI provides a 24-hour time coverage, the loops could be followed for 22 hours and a rise speed of $1.86\pm0.04$\,km\,s$^{-1}$ was measured. 

\begin{figure}[h]
\centerline{\includegraphics[width=.46\textwidth]{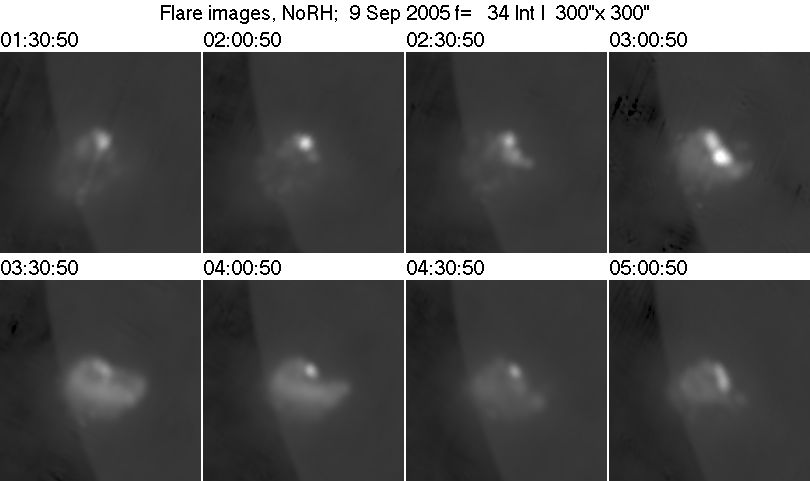}}
\caption{Development of loops after the X1.1 flare of September 9, 2005, at 03:07\,UT, observed at 34\,GHz. All images are displayed with the same intensity scale.}
\label{fig:Flare}
\end{figure}

From September 8 onward, TRACE was continuously pointed at AR 10808, and this gave us the opportunity to follow the loop evolution in the 171\,\AA\ and 1600\,\AA\ bands. A set of images is given in Figure~\ref{fig:EvolSep8}, together with SXI and \ha. In addition to the expansion of the loops and the 
changes in their form, it is obvious that the hot loop system decayed more slowly than the cold one. The loop expansion velocity, measured from the SXI images, was $\sim0.4$\,km/s, considerably smaller than that of the previous day. 

\begin{figure*}
\centerline{\includegraphics[width=.98\textwidth]{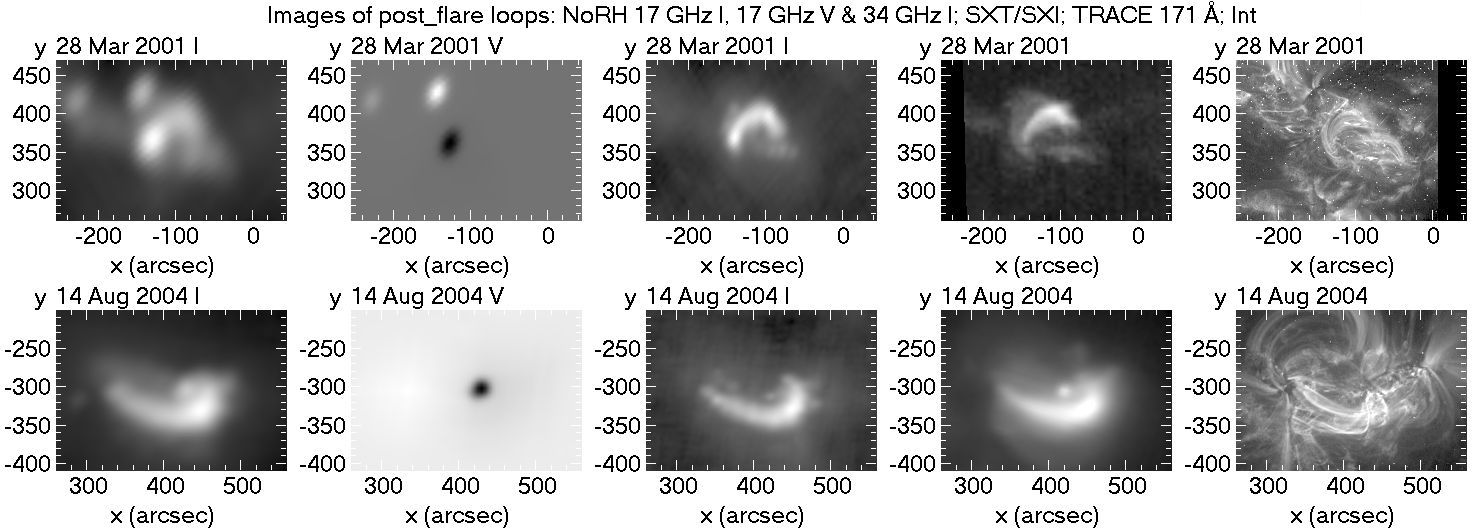}}
\caption{Images of loops in active regions 9393 (top row) and 10656 (bottom row). 17\,GHz Stokes I images are in the first column, 17\, GHz Stokes V in the second, 34\,GHz I in the third, Hinode/SXT (top) or GOES/SXI (bottom) in the fourth column and TRACE 171\,\AA\ in the fifth.}
\label{fig:other}
\end{figure*}

\begin{figure*}
\centerline{\includegraphics[width=.98\textwidth]{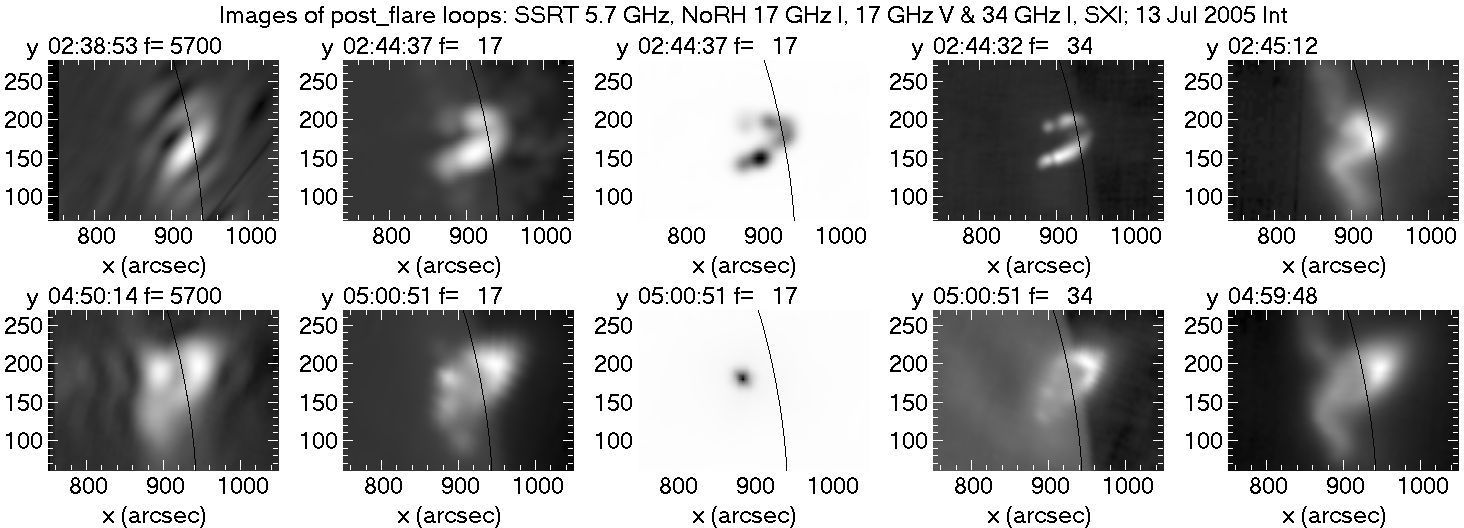}}
\caption{Images of loops in active region 10786 during the impulsive phase (top row) and during the post-flare phase (bottom row). Columns, from left to right, show SSRT images at 5.7\,GHz, 17\,GHz Stokes I images, 17\,GHz Stokes V, 34\,GHz Stokes I and GOES/SXI. The black arc marks the photospheric limb.}
\label{fig:Jul2005}
\end{figure*}

The development and subsequent decay of loops after each large flare is seen well in SXI and TRACE images. One case, with a loop system that developed after a flare that occurred  at 03:00\,UT on September 9, during the NoRH observations, is shown in Figure~\ref{fig:Flare}. 

As the flares occurred during Nobeyama night and the associated loops had a long duration, the NoRH images alone would give an impression of a single, long-lived, loop system. We note that the post-flare character of the loops explains some of their properties, such as their multi-thermal nature and their expansion. We further note that, in contrast to the post-flare loops, quiescent 171\,\AA\ loops (e.g. on either side of the large loop on September 7, Figure~\ref{fig:overview}) were not visible in the NoRH images. 

\section{Loops in other active regions}\label{sect:other}
\cite{2023ApJ...943..160F} presented 57 cases of high $T_b$ sunspot-associated emission at 34\,GHz, from 27 active regions in their table 1. In addition to the cases reported above, we found 11 more cases of post-flare loops in 8 active regions \citep[events  1, 2, 6, 8, 25, 26, 28, 29, 35, 55 and 57 in table 1 of][]{2023ApJ...943..160F}. With the exception of the June 13, 2005 event, these events were shorter and less impressive than the ones presented in the previous section of this article, yet we selected three of them that showed well-defined loop structures and we will present them here. 

In AR 9393 (top row in Figure~\ref{fig:other}) the E footpoint of the loop was strongly polarized, indicating gyrosynchrotron emission from the flare, but the rest of the loop was unpolarized, indicating thermal free-free emission. The only polarized emission in AR 10656 (bottom row in the figure) came from the sunspot-associated source; the loop showed no detectable polarization, indicating again free-free emission. 

For AR 10786 we give two image sets (Figure~\ref{fig:Jul2005}), one during the impulsive phase (top row) and one during the post-burst phase (bottom row), which demonstrate nicely the transition from polarized gyrosynchrotron to unpolarized thermal free-free emission.
In this case a loop morphology change was also observed: the nonthermal phase showed a loop with bright legs and a fainter looptop, while the thermal phase showed a nice loop with a bright looptop and cusp; see also a front-page figure of a White Paper by \citet{2009astro2010S..13B}. This particular event was fully covered by the NoRH, as well as by the SSRT and was also observed by RHESSI; however, a detailed study is outside the scope of this article.

The soft X-ray images are very similar to the NoRH images, having about the same resolution. The TRACE images show many more loops, thanks to their superior resolution of 0.5\arcsec. Moreover there are structural differences with respect to SXT or SXI, apparently due to the higher temperature of formation of the latter. In particular, in AR 9393 (top row of Figure~\ref{fig:other}) the TRACE image shows some features near the western leg of the loop, and the image of AR 1056 (bottom row) shows a horizontal loop below the main one, not visible in the other wavelength ranges.

\begin{figure}
\centerline{\includegraphics[width=.46\textwidth]{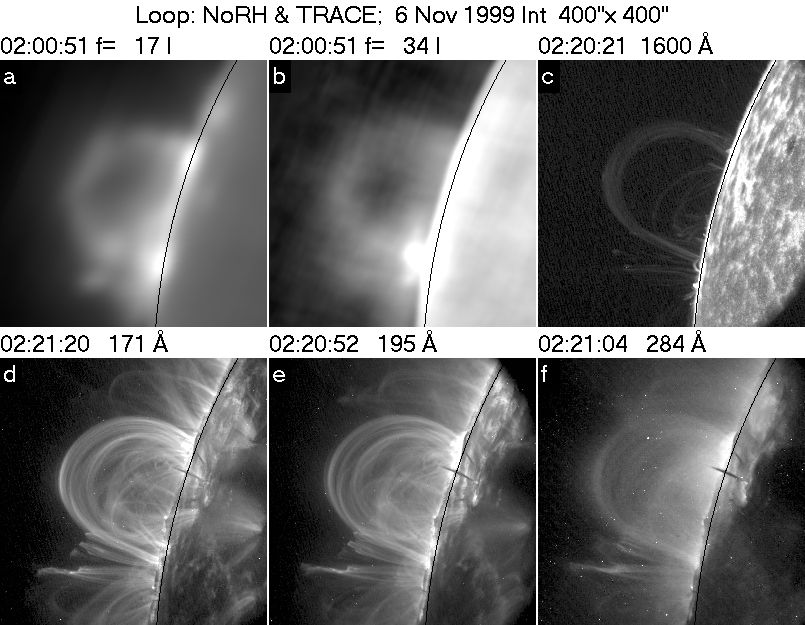}}
\caption{Images of the loop of November 6, 1999: (a) NoRH 17\,GHz, I; (b) NoRH 34\,GHz, I; (c)-(f) TRACE 1600\,\AA, 171\,\AA, 195\,\AA, and 284\,\AA.}
\label{fig:Quiescent}
\end{figure}

To the three loops taken from the work of \cite{2023ApJ...943..160F}, we added a spectacular loop system imaged by TRACE on November 6, 1999 and shown in figure 3c of \cite{2000ApJ...541.1059A}. This loop complex is probably seen face-on and was about 110\,Mm high and   120\,Mm wide (Figure~\ref{fig:Quiescent}). The images shown in the figure were obtained about 8 hours after the associated GOES M 3.0 flare (SOL1999-11-05T18:04:00) that occurred behind the limb, with the expanding loops well visible in EIT 195\,\AA\ images. An animatioon of the EIT images of the loop is available at https://zenodo.org/records/11357797. 

\section{Data analysis}
We adopt here that the NoRH radio emission is thermal free-free, as evidenced by the lack of strong circular polarization. Moreover, as the NoRH observations were obtained some hours after the flares, we would not expect that the post-flare loops would be filled with trapped energetic electrons. Energetic electrons could still be produced by reconnections near the cusp of the loops \citep{2022Natur.606..674F} and these might be associated with the noise storm observed at metric wavelengths in association with the events of September 2005 (Sect.~\ref{sect:10808}).

Moreover, our analysis is limited by instrumental issues, such as the limited resolution of the radio images and the lack of EUV data in a sufficient number of channels to compute the differential emission measure (DEM); the latter imposes to us to follow an isothermal approach, although it is well known and amply demonstrated in the previous sections that post-flare loops are multi-thermal.

\subsection{Computation of physical parameters}
Let us start from the expression for the brightness temperature, $T_b$ of an homogeneous, isothermal source with electron temperature $T_e$ and optical depth $\tau$:
\be
T_b=T_e(1-e^{-\tau})
\ee
with the optical depth given by:
\be
\tau=\int_0^L \frac{\xi}{f^2}\frac{N_e^2}{T_e^{3/2}} d\ell=\frac{\xi}{f^2 T_e^{3/2}} EM \label{eq:tau}
\ee
where $f$ is the observing frequency,
$\xi(f,N_e,T_e)$ is a slowly varying factor encompassing the ionization states and the Gaunt factor, and $EM$ is the emission measure: 
\be
EM=\int_0^L N_e^2 d\ell \label{eq:em}
\ee
where $N_e$ is the electron density and the integration is along the line of sight. $\xi$ was computed from \cite{2021ApJ...914...52F}; we note that equation (6.1) of \cite{1996ASSL..204.....Z} gives lower values by 15\%.

The above relations cannot be applied directly to our data, because we do not resolve individual loops. Still, we can measure with a great degree of accuracy the integral of $T_b$ in the direction $x$ perpendicular to the loops, to obtain the one-dimensional (1D) flux, $S_x$: 
\be
S_x=\int_0^X T_b(x) dx
\ee
We note that the 1D flux requires no correction for instrumental resolution, as the length of the loop is significantly larger than the NoRH beam. In our case it was measured along the approximate axis of symmetry of the loop, crossing its top.

Let us further assume that we have $N$ similar loops, each of width $w$. The 1D flux can then be expressed as:
\be
S_x= N w T_e (1-e^{-\tau}) \label{eq:S_x} 
\ee
If the structures are optically thin, their exact location and overlapping does not affect the result; in the opposite case the obscuration of the background loops must be taken into consideration. 

$N$ and $w$ may be estimated by visual inspection of high resolution EUV images. An additional constraint on $Nw$ could be set from the width of the loop group, $W$, measured at 34\,GHz; for example, you cannot pack more than $(W/w)^2$ loops in a region of width $W$, so that we can write:
\be
Nw^2<W^2 \label{eq:Nw}
\ee

Thus, if we can provide reasonable estimates for the temperature, $N$, and $w$, (\ref{eq:S_x}) allows us to compute $\tau$, and from that $EM$ and $N_e$ from observations at a single frequency: 
\bea
\tau&=&-\ln \left(1-\frac{S_x}{N w T_e}\right) \label{eq:tau_thick} \\
EM &=& \tau \frac{f^2 T_e^{3/2}}{\xi}  \label{eq:EM_tau} \\
N_{e}&\simeq&\sqrt{EM/w} \label{eq:EMw}
\eea

The situation is better if we have measurements at two frequencies, $f_1$ and $f_2$; applying (\ref{eq:S_x}), we obtain:
\be
\frac{S_x(f_1)}{S_x(f_2)}= \frac{1-e^{-\tau(f_1)}}{1-e^{-\tau(f_2)}} \label{eq:sxrat}
\ee
which, together with
\be
\frac{\tau(f_1)}{\tau(f_2)} =\frac{\xi(f_1)}{\xi(f_2)}\left(\frac{f_2}{f_1}\right)^2 \label{eq:taurat}
\ee 
can be solved for $\tau(f_1)$ and $\tau(f_2)$.

This approach is not valid if both frequencies are optically thin, because in this case (\ref{eq:sxrat}) gives simply:
\be
\frac{S_x(f_1)}{S_x(f_1)}= \frac{\tau(f_1)}{\tau(f_2)} \label{eq:sxrat2}
\ee
and the system (\ref{eq:sxrat}) and (\ref{eq:taurat}) is indeterminate. For the optically thin case we have a simpler relation for the optical depth:
\be
\tau=\frac{S_x}{N w T_e}
\ee
hence, in this case:
\bea
EM&\propto&\frac{\sqrt{T_e}}{Nw}\\ \label{eq:EMprop}
N_e&\propto&\frac{T_e^{1/4}}{\sqrt{N}w} \label{eq:Neprop}
\eea

\begin{table*}
\caption{Minimum $T_e$ and optical depth for the loops of July and September 7, 2005}\label{table:Te_min}
\begin{center}
\begin{tabular}{cc|ccc|ccc}
\hline
   Date     &   UT &      $S_{x34}$  &      $S_{x5.7}$ &$W_{34}$ &$\tau_{34}$&$\tau_{5.7}$&$T_{e_{min}}$\\ 
            &      &MK$\times$\arcsec&MK$\times$\arcsec&\arcsec\ &           &            &    MK     \\
\hline
Jul 13, 2005& 04:00& 5.27 & 41.4 & 28 & 0.135 & 4.68 &  1.49 \\
            & 04:30& 3.85 & 44.3 & 28 & 0.086 & 2.99 &  1.67 \\
            & 05:00& 2.12 & 38.5 & 28 & 0.044 & 1.53 &  1.76 \\
Sep  7, 2005& 02:30& 1.56 & 22.5 & 36 & 0.064 & 2.20 &  0.70 \\
\hline
\end{tabular}
\end{center}
\end{table*}

In the optically thick case, obscuration effects change (\ref{eq:Nw}) to
\be
Nw<W
\ee
in which case, combining with (\ref{eq:S_x}), we obtain a lower limit for the loop temperature:
\be
T_e>\frac{S_x}{W(1-e^{-\tau})} \label{eq:TeC}
\ee
Note that this lower limit is equal to $T_e$ for a single loop of width $w=W$

Finally, once we have the loop density and temperature, we can estimate a lower limit to the magnetic field required for the confinement of the plasma, by equating the plasma pressure, $2N_ekT_e$, to the magnetic pressure, $B^2/8\pi$.

\subsection{Application to the measurements}

\subsubsection{Optical thickness at the NoRH frequencies}
The ratio of the brightness temperature at two frequencies in the optically thin case is equal to the ratio of the optical depth. Putting numerical values for the NoRH frequencies, we obtain:
\be
 \frac{T_b(34\,\mbox{GHz})}{T_b(17\,\mbox{GHz})}=0.238 \label{eq:TbRatVal}
\ee
In the optically thick case, the ratio will be grater than the value given in (\ref{eq:TbRatVal}), approaching unity for $\tau\gg 1$.

A prerequisite for the application of the above criterion is that the images at the two frequencies have the same spatial resolution. To that end, we selected a pair of good images for each day during the September 2005 period, removed the quiet sun background by subtracting the azimuthal average of the brightness temperature, masked the emission from the active region when necessary and smoothed the 34\,GHz images to the 17\,GHz resolution. Subsequently we made scatter plots of $T_b$ at the two frequencies and measured the slope. The advantage of this method is that it provides a point-by-point comparison and avoids issues such as the exact value of the background. For all days we obtained slopes near the value predicted by (\ref{eq:TbRatVal}), which provides a strong indication that we are in the optically thin regime at both NoRH frequencies. 

The above method could not be applied to the SSRT images, because their resolution was not high enough. We note, however, that if the 5.7\,GHz emission were optically thin, $T_b$ should have been a factor of 9.5 higher than that at 17\,GHz. As an example, for September 7, 2005 this translates to a peak brightness of $1.4\times10^6$\,K; as this value is well above the observed $T_b$ of $\sim5\times10^5$\,K, we conclude that the 5.7\,GHz emission from the loops was optically thick.

\begin{table*}
\caption{Emission measure, density and magnetic field  for the loops of Table~\ref{table:Te_min}}\label{table:EM_den}
\begin{center}
\begin{tabular}{cc|ccc|ccc|ccc}
\hline
   Date     &   UT &\multicolumn{3}{c}{EM ($10^{29}$\,cm$^{-5}$)}&\multicolumn{3}{c}{$N_e$ ($10^{10}$\,cm$^{-3}$)}&\multicolumn{3}{c}{$B$ (G)}\\
            &      &$T_{e_{min}}$&$2\times10^6$\,K &$5\times10^6$\,K&$T_{e_{min}}$&$2\times10^6$\,K &$5\times10^6$\,K&$T_{e_{min}}$&$2\times10^6$\,K &$5\times10^6$\,K\\
\hline
Jul 13, 2005& 04:00& 17.6& 26.8& 99.8& 9.0& 11.1& 21.4& 18.2& 39.3& 86.2 \\
            & 04:30& 13.2& 17.1& 63.7& 7.8& ~8.8& 17.1& 16.9& 35.1& 77.1 \\
            & 05:00& ~7.3& ~8.8& 32.6& 5.8& ~6.3& 12.2& 14.6& 29.7& 65.2 \\
Sep  7, 2005& 02:30& ~3.2& 14.4& 53.5& 3.9& ~8.1& 11.6& 12.9& 33.6& 73.8 \\
\hline
\end{tabular}
\end{center}
\end{table*}

\subsubsection{Estimate of the electron temperature}
As noted above, we can obtain lower limits of $T_e$ from the optically thick SSRT images. From the images during the September 2005 period (Figure~\ref{fig:overview}), only that of September 7 is usable because for  other days the images were contaminated by the sunspot-associated emission. We also have three image sets during the July 13, 2005 post-flare loops, in which SSRT images were close enough in time to NoRH images.

In a first step the optical depth at 34\,GHz and 5.7\,GHz was computed from (\ref{eq:sxrat}) -- (\ref{eq:taurat}) and the corresponding $S_x$ values. Subsequently $T_{e_{min}}$ was computed from (\ref{eq:TeC}), using the width, $W$, measured from the 34\,GHz images, after correction for the instrumental resolution. The results are given in Table~\ref{table:Te_min}.

We note that for the July loops the optical depth decreases with time, as expected during the decay phase, whereas $T_{e_{min}}$ increases. For this period the lower limit of the temperature is above $10^6$\,K, whereas it is below that for the September loops. In all cases the optical depth at 34\,GHz is comfortably low, whereas it is above 2 at 5.7\,GHz, except for one case.

We still need an upper limit of $T_e$. To that end, we computed the EM from (\ref{eq:EM_tau}), the density from (\ref{eq:EMw}) and the magnetic field for $w=3$\arcsec, for three values of $T_e$: the minimum, 2\,MK and 5\,MK. They are all given in Table~\ref{table:EM_den}.

Note that the emission measure depends only on  the assumed $T_e$ ($\tau$ is derived from the $S_x$ ratio), and not on the number of loops, $N$, the width of the loops, $w$ or the width of the loop system, $W$. As expected from (\ref{eq:EM_tau}), it increases with $T_e$. On the other hand, $N_e$ has values above $10^{10}$, exceeding $10^{11}$ for the higher $T_e$.

The values of $B$ for $T_e=5\times10^6$\,K are, likely, too high, compared with previous estimates from polarization inversion (this method is probably the most reliable of all) around the height of 100\,Mm \citep[see table 1 of][which gives values between 10 and 20\,G]{2021FrASS...7...77A}; see also figure\,1 in \citet{2023ApJ...943..160F}. Considering this and the high density values derived for $T_e=5\times10^6$\,K, we will adopt a temperature of $2\times10^6$\,K, as a reasonable estimate 

We note that the adopted value of $T_e$ is above the lower limits provided by the observed $T_b$ and is close to the the formation temperature of the EUV emission ($10^6$\,K), whereas the soft X-ray emission is probably produced by hotter plasma. It is possible that, at the adopted $T_e$, the loops are more diffuse and less numerous than in the 171\,\AA\ images, in which case $w$ would be larger and the derived $N_e$ would be lower, as it goes like $\sqrt{(1/w)}$.

\subsubsection{Emission measure and electron density}
\label{sect:compute}
We had no SSRT images for the other loop systems, thus for the estimates of the physical parameters we used the relations (\ref{eq:tau_thick}) to (\ref{eq:EMw}) to compute $\tau$, EM and $N_e$, which require estimates of $T_e$, $N$ and $w$. For the temperature we used a flat value of 2\,MK, as discussed in the previous section; in any case, due to the weak dependence of the computed $N_e$ on the assumed $T_e$ ($N_e\propto T_e^{1/4}$), the density would increase only by a factor of 1.5 if the temperature went up from $2\times10^6$\,K to $10^7$\,K and would decrease by a factor of 1.2 if $T_e$ went down to $10^6$\,K. We counted more than a dozen loops on the 171\,\AA\ image of September 8, 2005 and took $N=20$ for the September 2005 loops, except for the long arcade of September 10, where we used $N=50$. We measured a width of $\sim3$\arcsec\ on the TRACE images and used that value. We used he same values of $N$ and $w$ for the other loops of our collection, except for the event of November 6, 1999 which consisted of a multitude of loops, for which we set $N=50$.

\begin{table*}
\caption{Loop measurements at 17\,GHz and derived parameters}\label{parms_17}
\begin{center}
\begin{tabular}{lcc|cccccc}
\hline
Date  &Time &$\Delta t$
&1D flux                   &     1D EM              &     EM              &   $N_e$            & $B$ &$\tau$\\
      & UT  &hours&10$^6$\,K$\times$\arcsec\ &10$^{31}$\,cm$^{-4}$ &10$^{29}$\,cm$^{-5}$ &$10^{10}$\,cm$^{-3}$&  G  &\\
\hline
   Sep ~\,7, 2005 &00:00& 3.5 & 8.92 &1.97 &3.28 &3.76 &22.9 &0.073 \\
                  &06:00& 8.5 & 3.34 &0.72 &1.19 &2.27 &17.8 &0.026 \\
\hline
   Sep ~\,7, 2005 &22:50& 5.6 & 9.66 &2.14 &3.56 &3.92 &23.3 &0.079 \\
   Sep ~\,8, 2005 &04:30&11.2 & 3.23 &0.69 &1.15 &2.23 &17.6 &0.026 \\
\hline
   Sep ~\,8, 2005 &22:55& 2.0 & 18.8 &4.36 &7.27 &5.60 &27.9 &0.161 \\
   Sep ~\,9, 2005 &03:35& 0.9 & 21.9 &5.14 &8.56 &6.08 &29.1 &0.189 \\
\hline
   Sep ~\,9, 2005 &22:50& 3.6 & 43.8 &10.1 &6.71 &5.38 &27.3 &0.148 \\
     Sep 10, 2005 &06:25&11.2 &  3.3 & 0.7 &0.46 &1.41 &14.0 &0.010 \\
\hline
   Mar 28, 2001   &00:00& 3.5 & 12.7 &2.84 &4.74 &4.52 &25.1 &0.105 \\
\hline
   Aug 14, 2004   &01:00& 1.1 & 10.2 &2.26 &3.76 &4.03 &23.7 &0.083 \\
\hline
Jul \,\,13, 2005  &04:00& 1.3 & 21.5 &5.03 &8.39 &6.02 &28.9 &0.185 \\
                  &04:30& 1.8 & 14.7 &3.33 &5.56 &4.90 &26.1 &0.123 \\
                  &05:00& 2.3 & 10.3 &2.28 &3.81 &4.05 &23.7 &0.084 \\
                  &05:40& 3.0 &  7.5 &1.63 &2.72 &3.43 &21.8 &0.060 \\
\hline
Nov \,\,\,6, 1999 &02:06& 7.5 &  1.1 &0.22 &0.14 &0.80 & 10.5 &0.003 \\
\hline
\end{tabular}
\end{center}
\end{table*}

\begin{table*}
\caption{Loop measurements at 34\,GHz and derived parameters}\label{parms_34}
\begin{center}
\begin{tabular}{lcc|cccccc}
\hline
Date  &Time &$\Delta t$
&1D flux                   &     1D EM              &     EM              &   $N_e$            & $B$ &$\tau$\\
      & UT  &hours&10$^6$\,K$\times$\arcsec\ &10$^{31}$\,cm$^{-4}$ &10$^{29}$\,cm$^{-5}$ &$10^{10}$\,cm$^{-3}$&  G  &\\
\hline
   Sep ~\,7, 2005 &00:00& 3.5 & 1.39 &1.24 &2.06 &2.99 &20.4 &0.011 \\
                  &06:00& 8.5 & 0.92 &0.81 &1.36 &2.42 &18.3 &0.007 \\
\hline
   Sep ~\,7, 2005 &22:50& 5.6 & 2.58 &2.31 &3.85 &4.08 &23.8 &0.020 \\
   Sep ~\,8, 2005 &04:30&11.2 & 1.05 &0.94 &1.56 &2.60 &19.0 &0.008 \\
\hline
   Sep ~\,8, 2005 &22:55& 2.0 & 5.46 &4.95 &8.25 &5.97 &28.8 &0.044 \\
   Sep ~\,9, 2005 &03:35& 0.9 & 5.90 &5.36 &8.93 &6.21 &29.4 &0.047 \\
\hline
   Sep ~\,9, 2005 &22:50& 3.6 & 10.5 &9.44 &6.29 &5.21 &26.9 &0.033 \\
     Sep 10, 2005 &06:25&11.2 &  0.6 &0.54 &0.36 &1.24 &13.1 &0.002 \\
\hline
   Mar 28, 2001   &00:00& 3.5 & 2.77 &2.48 &4.13 &4.23 &24.2 &0.022 \\
\hline
   Aug 14, 2004   &01:00& 1.1 & 3.26 &2.93 &4.88 &4.59 &25.3 &0.026 \\
\hline
Jul \,\,13, 2005  &04:00& 1.3 & 5.27 &4.77 &7.95 &5.86 &28.5 &0.042 \\
                  &04:30& 1.8 & 3.85 &3.46 &5.77 &4.99 &26.3 &0.031 \\
                  &05:00& 2.3 & 2.12 &1.89 &3.15 &3.69 &22.6 &0.017 \\
                  &05:40& 3.0 & 2.34 &2.09 &3.48 &3.88 &23.2 &0.018 \\
\hline
Nov \,\,\,6, 1999 &02:06& 7.5 & 0.31 &0.27 &0.18 &0.88 & 11.1 &0.001 \\
\hline
\end{tabular}
\end{center}
\end{table*}

In Tables~\ref{parms_17} and \ref{parms_34} we give our measurements and the results of our computations for 17\,GHz and 34\,GHz respectively, for all loop systems, including the ones in Table~\ref{table:EM_den}; the loops are ordered as presented in Sections \ref{sect:10808} and \ref{sect:other}. As mentioned previously, the measurements refer to the top of the loops

The Tables give the measured 1D flux, the computed 1D emission measure, the emission measure and the density of an individual loop; we also provide our estimate of the magnetic field, as well as the optical depth of an individual loop as a check of the optical thickness. $\Delta t$ is the time elapsed since the associated flare. For the September 2005 loops we provide two entries for each observing day, one near the beginning and one near the end of the observation or of the period that the loop could be followed, in order to give an idea of the decay of the emission; the exception is the entry for September 9 at 03:30\,UT, which refers to the start of the new loop shown in Figure~\ref{fig:Flare}. For the July 13, 2005 loops we give four values during the post-flare phase and one value for the other loops. Results for other values of $w$, $N$ and $T_e$ can be obtained by scaling the table values according to equations (14) to (\ref{eq:Neprop}). For example, if the loops were more diffuse and less numerous, $EM$ would not be affected (eq.~15) and $N_e$ would be lower, as discussed in the last paragraph of the previous section.

We first note the similarity of the values of $EM$ and $N_e$ obtained from data at 17\,GHz and at 34\,GHz for the same date and time, which adds confidence to our approach. We also note that the computed values of $\tau$ verify that the loops are optically thin at both NoRH frequencies. 

There are large variations among the values of the derived physical parameters, due to both intrinsic differences and different evolution phases in the course of the each event. The emission measure is in the range of $\sim0.2\times10^{29}$\,cm$^{-5}$ to $\sim9\times10^{29}$\,cm$^{-5}$, while $N_e$ ranges from $\sim10^{10}$\,cm$^{-3}$ to $\sim6\times10^{10}$\,cm$^{-3}$. The derived densities are several times above those deduced from observations in other wavelength regimes (up to $\sim10^{10}$\,cm$^{-3}$, see Section~\ref{sect:intro}); they are likely even higher closer to the flare peaks. The density values are about two orders of magnitude above the upper limit suggested by the NRH data (Section~\ref{sect:NRH_thermal}) for the 432\,MHz source, whose apparent location was near the top of the SXI loops (Figure~\ref{fig:NRH_SXI}); thus, if the metric source was indeed associated with the loops, its emission should have originated in a less dense region, possibly in the sheath of the loops. 

Compared to the values in Table~\ref{table:EM_den} for the July 13, 2005 loops, computed for $T_e=2\times10^6$\,K, the density values given here are lower by almost a factor of 2, apparently due to the different methods of computation. It could be reduced by lowering the value of $w$ to 1\arcsec\ or less but, in any case, this provides an estimate of the accuracy of our computations.

The time variation of the 1D flux and the emission measure provide evidence about the duration of the post-flare loop phase; for example, during the September 10 observations, the emission dropped by more than a factor of 10 in 7.5 hours. A least square fit to an exponential decay gave a characteristic time of 2.9\,h for the 1D flux; as $N_{e0}\propto\sqrt S_x$ and $S_x$ depends weakly on the temperature, the decay time for the density is twice that value. For September 7 and 8, the 1D flux decay times were 5\,h and 4.3\,h, respectively, whereas on September 9 the loop emission could not be reliably isolated from the  active region emission for a long enough interval to compute a meaningful decay time. Overall, mm-$\lambda$ post-flare loop emission was detectable up to at least 11\,h after the associated flare.

Finally, the estimated value of the magnetic field required to confine the loop plasma is in the range of 10-30\,G. Interestingly, the estimates of the longitudinal component of the magnetic field, obtained from the polarization of the thermal free-free emission observed with the NoRH and SSRT  (Section~\ref{sect:overview}) were in the same range.

\section{Summary}\label{sect:conclude}
We reported the detection of thermal emission from post-flare loops at 34\,GHz  and  computed their electron density from NoRH observations at both 17\,GHz and 34\,GHz. In addition to these two frequencies, our analysis included data from other instruments, covering a broad wavelength range and a temperature range from the chromosphere to the corona. No quiescent loops were detected by the NoRH, apparently due to their lower density.

Morphologically our loops conformed to classic post-flare loops described in the literature: they lasted several hours, were seen from \ha\ to soft X-rays and displayed  an apparent expansion. Their projected height was up to 110\,Mm and the distance between their footpoints went up to 120\, Mm. In two cases (September 8 and 9, 2005, see Figure~\ref{fig:overview}), long arcades of post-flare loops were detected. 

We recalibrated the 34\,GHz images using the disk center $T_b$ deduced from the analysis of ALMA full-disk images by \cite{2023A&A...670C...5A}, which gave a correction factor of 0.8.The absence of strong circular polarization shows that the emission was thermal free-free. By comparing the brightness of the loops at the two NoRH frequencies, we concluded that the emission was optically thin at both; at the longer (5.7\,GHz) wavelength of SSRT the loops were optically thick. 

Although the loops are multi-temperature structures, as evidenced from their appearance in many wavelength ranges, we adopted an isothermal assumption, being limited to observations at only two radio frequencies. An alternative approach would be to compute the brightness temperature from the differential emission measure deduced from EUV data \citep[c.f.][]{2013PASJ...65S...8A}, but our events were before the SDO era.

As the NoRH resolution is not sufficient to resolve individual loop threads, we computed observable parameters for a bundle of similar, unresolved, optically thin loops, in terms of the emission measure and the temperature of a single loop. In particular, we computed the 1D flux, $S_x$, which is the integral of $T_b$ across the loop and is directly measurable from the images; moreover, $S_x$ does not depend upon the instrumental resolution, provided that the loop length is sufficiently grater that the beam size. The 1D emission measure, $EM_x$, which we defined as the integral of the emission measure across the loop can be computed from $S_x$, under a reasonable estimate of the electron temperature. We further computed the emission measure and density of an individual loop, assuming that we have $N$ similar loops of width $w$ in the bundle of unresolved loops. 

Lower limits of the electron temperature of about $5\times10^5$ were provided by measurements of the observed brightness temperature. Better estimates of $T_e$ were obtained in a few cases where we had simultaneous SSRT images at 5.7\,GHz, where the optical thickness of the emission was higher; based on that analysis we adopted $T_e=2\times10^6$\,K, not far from the formation temperature of 171\,\AA\ band observed by TRACE. We measured 3\arcsec\ as the average width of loops in TRACE images and we used this value. We counted about a dozen loops and set $N=20$ for most cases, except for a long arcade and a loop of many threads for which we we used $N=50$.

With these parameters we obtained values of 0.3--$10\times10^{31}$\,cm$^{-4}$ for the 1D emission measure, 0.2--$9\times10^{29}$\,cm$^{-5}$ for the emission measure and 1--$6\times10^{10}$\,cm$^{-3}$ for the density of individual loops. These values refer to the top of the loops and vary significantly from one loop to another and in the course of the evolution of the same loop system. The deduced density depends weakly on the assumed electron temperature ($N_e\propto T_e^{1/4}$) and goes like $(w\sqrt N)^{-1}$. Similar results were obtained for both NoRH frequencies, while the combination of NoRH and SSRT images for the loop of July 13, 2005 gave higher values by about a factor of two.

By equating the magnetic pressure to the gas pressure we obtained estimates of 10--30\,G for the magnetic field. Interestingly, this range of values is compatible with the magnetic field estimates deduced from the polarization of the radio emission.
 
We did not study the decay of the loops in detail, but we noted by visual inspection of the images that the hot loops decayed more slowly than the cold ones. Least square fits to an exponential decay for three loop systems gave characteristic times from 2.9\,h to 5\,h for the 1D flux and twice as much for the density ($N_{e}\propto\sqrt S_x$). These values are noticeably shorter than the conductive cooling time for $L\sim 100$\,Mm, $n_e\sim10^{10}$\,cm$^{-3}$, and $T_e\sim1-5$\,MK, but longer than the radiative cooling time for the same range of parameters. Whether these values reflect the evolution of the individual loops or the phenomenon as a whole, they require a sustained loop heating due to extended energy release. 

In the metric wavelength range, NRH images showed noise storm emission above the loops observed in September 2005. In one case a low brightness source at 432\,MHz, which could be thermal, coincided with the top of the loop system above the limb. The plasma dispersion relation puts an upper density limit of two orders of magnitude below the value derived from the NoRH data, so that if an association existed indeed, the metric emission must have come from the outer, less dense, part of the loop.

\section{Discussion and conclusions}\label{sect:discuss}
Our range of densities is above that reported from EUV observations, which go up to about $10^{10}$\,cm$^{-3}$ (Section~\ref{sect:intro}). In particular we note that for the November 6, 1999 loop the density given here is one order of magnitude above the density computed from the intensity ratio of TRACE 171\,\AA\ and 195\,\AA\ images by \cite{2000ApJ...541.1059A} for the same loop. As the mechanism of radio emission is well understood, we believe that our results, in spite of the limited resolution of our data, are more reliable than those in the EUV and in the X-rays, which suffer from inherent uncertainties concerning abundances and non-LTE excitation/ionization equilibria. Alternatively, the difference could be attributed to the fact that different regions of the loops are sampled in the two spectral domains. 

As almost all cases studied occurred in active regions containing sunspots with extremely high magnetic field, an interesting question is if there is a physical association between the two; as a matter of fact, post-flare loops were detected in 9 out of the 27 regionsstudied by \cite{2023ApJ...943..160F}. Of course these were active regions with highly complex magnetic configurations and thus more likely to produce large flares, so that we cannot give a definitive answer to that question, but only provide a hint. We note that the estimated $B$ values are all around the upper bound suggested by figure 1 of  \cite{2023ApJ...943..160F} and by Table 1 of \cite{2021FrASS...7...77A}. This might imply that the presence of strongest magnetic field at the photosphere/base of the corona is a prerequisite of having sufficiently strong magnetic field at the tops of such giant dense loops to keep them magnetically confined.

This work demonstrates the potential of radio observations for the study of post-flare loops. Spatial resolution is a limitation, and this could be overcome with ALMA \citep{2023FrASS..1038626A}; however, ALMA solar observing is scarce and that of prominences even more so \citep{2022FrASS...9.3707H}. Important information can be obtained from multi-frequency radio observations in a wavelength range covering both optically thin emission at shorter $\lambda$ and optically thick emission at longer $\lambda$. In principle, such observations could provide a differential emission measure, relaxing the isothermal assumption. Modern imaging spectroscopy instruments, such as the Extended Owens Valley Solar array (EOVSA), the Siberian Radioheliograph (SRH) and the Mingantu Ultrawide Spectral Radioheliograph (MUSER) could provide such opportunities in the future.

\medskip
GF is supported by NSF grants AGS-2121632 and AST-2206424  
and NASA grant 80NSSC23K0090. The authors express their gratitude to the colleagues and staff of the NoRH, TRACE, GOES, GOES/SXI, Hinode/SXT, SOHO/EIT, WIND/ WAVES, the  Kanzelhoehe solar observatory and the NRH for making their data publicly available. We also wish to thank K. Shibasaki, M. Shimojo and T. Bastian for interesting discussions.

\bibliographystyle{aasjournal}
\bibliography{References,More,fleishman}

\end{document}

%% file: definitions.tex
\newcommand{\pr}{^\prime}
\newcommand{\pc}{\tl{pc}}
\newcommand{\kpc}{\tl{kpc}}
\newcommand{\Mpc}{\tl{Mpc}}
\newcommand{\ha}{H$\alpha$}

\newcommand{\vect}[1]{\mbox{\bf #1}}

\newcommand{\dotp}{\!\cdot\!}
\newcommand{\crs}{\!\times\!}

\newcommand{\be}{\begin{equation}}
\newcommand{\ee}{\end{equation}}
\newcommand{\bea}{\begin{eqnarray}}
\newcommand{\eea}{\end{eqnarray}}
\newcommand{\beas}{\begin{eqnarray*}}
\newcommand{\eeas}{\end{eqnarray*}}

\newcommand{\ddt}[1]{\frac{d{ #1}}{dt}}
\newcommand{\ddtt}[1]{\frac{d^2{ #1}}{dt^2}}
\newcommand{\ddr}[1]{\frac{d{ #1}}{dr}}
\newcommand{\ddx}[1]{\frac{d{ #1}}{dx}}
\newcommand{\prt}[1]{\frac{\partial{ #1}}{\partial t}}
\newcommand{\prtc}{\left. \frac{\partial f}{\partial t} \right|_c}
\newcommand{\prx}[1]{\frac{\partial{ #1}}{\partial \vect{x}}}
\newcommand{\prv}[1]{\frac{\partial{ #1}}{\partial \vect{v}}}
\newcommand{\prxi}[1]{\frac{\partial{ #1}}{\partial \vect{x}_i}}
\newcommand{\prvi}[1]{\frac{\partial{ #1}}{\partial \vect{v}_i}}
\newcommand{\prxo}[1]{\frac{\partial{ #1}}{\partial \vect{x}_1}}
\newcommand{\prvo}[1]{\frac{\partial{ #1}}{\partial \vect{v}_1}}

\newcommand{\prxx}[1]{\frac{\partial{ #1}}{\partial x}}
\newcommand{\pryy}[1]{\frac{\partial{ #1}}{\partial y}}
\newcommand{\przz}[1]{\frac{\partial{ #1}}{\partial z}}
\newcommand{\prrr}[1]{\frac{\partial{ #1}}{\partial r}}
\newcommand{\prth}[1]{\frac{\partial{ #1}}{\partial \vartheta}}
\newcommand{\prph}[1]{\frac{\partial{ #1}}{\partial \varphi}}

\newcommand{\vx}{\vect{x}}
\newcommand{\vz}{\vect{z}}
\newcommand{\vk}{\vect{k}}
\newcommand{\vs}{\vect{s}}
\newcommand{\vv}{\vect{v}}
\newcommand{\vV}{\vect{V}}
\newcommand{\vb}{\vect{B}}
\newcommand{\vB}{\vect{B}}
\newcommand{\ve}{\vect{E}}
\newcommand{\vE}{\vect{E}}
\newcommand{\vj}{\vect{J}}
\newcommand{\vJ}{\vect{J}}
\newcommand{\va}{\vect{a}}
\newcommand{\vpar}{\vect{v}_\parallel}
\newcommand{\vper}{\vect{v}_\perp}
\newcommand{\eper}{\vect{E}_\perp}
\newcommand{\ampE}{\tilde{\vE}_1}
\newcommand{\ampB}{\tilde{\vB}_1}
\newcommand{\amp}[1]{\tilde{#1}}
\newcommand{\skx}{\mbox{\scriptsize{\bf k}}\cdot\mbox{\scriptsize{\bf x}}}

\newcommand{\bu}{\hat{\vect{b}}}
\newcommand{\vu}{\hat{\vect{n}}}
\newcommand{\ru}{\hat{\vect{r}}}
\newcommand{\xu}{\hat{\vect{x}}}
\newcommand{\yu}{\hat{\vect{y}}}
\newcommand{\zu}{\hat{\vect{z}}}
\newcommand{\tu}{\hat{\mbox{\boldmath $\vartheta$}}}

\newcommand{\vbc}{\frac{\vv\times\vect{B}}{c}}
\newcommand{\Vbc}{\frac{\vV\times\vect{B}}{c}}
\newcommand{\jbc}{\frac{\vj\times\vect{B}}{c}}

\newcommand{\dv}{\nabla\dotp}
\newcommand{\rot}{\nabla\crs}
\newcommand{\Lap}{\nabla^2}

\newcommand{\qm}{\frac{q}{m}}
\newcommand{\inv}[1]{\frac{1}{ #1}}
\newcommand{\half}{\frac{1}{2}}
\newcommand{\opj}{\omega_{pj}}
\newcommand{\unten}{\mbox{\rm I}}